\documentclass[amsfonts, amsmath, amssymb, aps, pra, twocolumn, a4paper, superscriptaddress, floatfix]{revtex4-2}

\usepackage{physics}
\usepackage{changes}

\begin{document}
\title{Efficient option pricing with unary-based photonic computing chip and generative adversarial learning}

\author{Hui Zhang}
\affiliation{Institute of Quantum Technologies (IQT), The Hong Kong Polytechnic University, Hong Kong}
\affiliation{Quantum Science and Engineering Centre (QSec), Nanyang Technological University, Singapore}
\author{Lingxiao Wan}
\affiliation{Quantum Science and Engineering Centre (QSec), Nanyang Technological University, Singapore}
\author{Sergi Ramos-Calderer}
\affiliation{Departament de Fisica Quantica i Astrofisica and Institut de Ciencies del Cosmos (ICCUB), Universitat de Barcelona, Barcelona, Spain}
\affiliation{Quantum Research Centre, Technology Innovation Institute, Abu Dhabi, UAE}
\author{Yuancheng Zhan}
\affiliation{Quantum Science and Engineering Centre (QSec), Nanyang Technological University, Singapore}
\author{Wai-Keong Mok}
\affiliation{Centre for Quantum Technologies, National University of Singapore, Singapore}
\author{Hong Cai}
\affiliation{Institute of Microelectronics, A*STAR (Agency for Science, Technology and Research), Singapore}
\author{Feng Gao}
\affiliation{Advanced Micro Foundry, 11 Science Park Rd, Singapore}
\author{Xianshu Luo}
\affiliation{Advanced Micro Foundry, 11 Science Park Rd, Singapore}
\author{Guo-Qiang Lo}
\affiliation{Advanced Micro Foundry, 11 Science Park Rd, Singapore}
\author{Leong Chuan Kwek}
\email[Corresponding Author: ]{cqtklc@gmail.com (L.C.K), cqtjil@nus.edu.sg (J.I.L), eaqliu@ntu.edu.sg (A.Q.L)}
\affiliation{Quantum Science and Engineering Centre (QSec), Nanyang Technological University, Singapore}
\affiliation{Centre for Quantum Technologies, National University of Singapore, Singapore}
\affiliation{National Institute of Education, Nanyang Technological University, Singapore}
\author{Jos\'{e} Ignacio Latorre}
\email[Corresponding Author: ]{cqtklc@gmail.com (L.C.K), cqtjil@nus.edu.sg (J.I.L), eaqliu@ntu.edu.sg (A.Q.L)}
\affiliation{Departament de Fisica Quantica i Astrofisica and Institut de Ciencies del Cosmos (ICCUB), Universitat de Barcelona, Barcelona, Spain}
\affiliation{Quantum Research Centre, Technology Innovation Institute, Abu Dhabi, UAE}
\affiliation{Centre for Quantum Technologies, National University of Singapore, Singapore}
\author{Ai Qun Liu}
\email[Corresponding Author: ]{cqtklc@gmail.com (L.C.K), cqtjil@nus.edu.sg (J.I.L), eaqliu@ntu.edu.sg (A.Q.L)}
\affiliation{Institute of Quantum Technologies (IQT), The Hong Kong Polytechnic University, Hong Kong}
\affiliation{Quantum Science and Engineering Centre (QSec), Nanyang Technological University, Singapore}

\begin{abstract}
In the modern financial industry system, the structure of products has become more and more complex, and the bottleneck constraint of classical computing power has already restricted the development of the financial industry. Here, we present a photonic chip that implements the unary approach to European option pricing, in combination with the quantum amplitude estimation algorithm, to achieve a quadratic speedup compared to classical Monte Carlo methods. The circuit consists of three modules: a module loading the distribution of asset prices, a module computing the expected payoff, and a module performing the quantum amplitude estimation algorithm to introduce speed-ups. In the distribution module, a generative adversarial network is embedded for efficient learning and loading of asset distributions, which precisely capture the market trends. This work is a step forward in the development of specialized photonic processors for applications in finance, with the potential to improve the efficiency and quality of financial services.
\end{abstract}

\maketitle

\section{Introduction}
The pricing of financial derivatives is a prominent problem that requires extensive computational resources, as the stochastic nature of the underlying assets requires precise modeling. One of the typical financial derivatives is the option, which is a contract that allows the holder to buy or sell assets at a pre-established price (strike) at or before a specified date (maturity date). The payoff of an option relies heavily on the stochastic evolution of asset price. The traditional option pricing model, Black-Scholes-Merton model (BSM)~\cite{black2019pricing} usually oversimplifies market dynamics, which limits its practical application to real-life scenarios. As such, numerical methods such as the Monte Carlo method are typically employed for handling more realistic stochastic fluctuations. However, Monte Carlo method requires extensive computation resources and is slow to predict complicated options. Reducing the computational resources required for models and speeding up option pricing could have significant implications for the financial industry.

Recently, quantum algorithms have shown promise in facilitating computationally-hard financial problems like trading, portfolio optimization, and risk profiling~\cite{orus2019quantum, herman2022survey}, and specifically the quantum amplitude amplification can accelerate the option pricing with quadratic speedups~\cite{brassard2002quantum, montanaro2015quantum, rebentrost2018quantum, woerner2019quantum, focardi2020quantum, ramos2021quantum, bova2021commercial}. The unique advantages of quantum algorithms will make up for the shortcomings of classical algorithms to a certain extent, enabling massive high-speed data services in the financial industry. However, current experimental demonstrations using binary approaches and standard quantum circuit models on superconducting devices~\cite{stamatopoulos2020option} require dense chip connections and high gate fidelity, making it difficult for practical applications in the near future without a universal quantum computer~\cite{preskill2018quantum, bharti2022noisy}. In addition, superconducting devices require bulky, energy-intensive, and expensive peripherals like cooling systems, making industrial-scale applications poor prospects.

\begin{figure*}[t]
\centering
\includegraphics[width=0.69\textwidth]{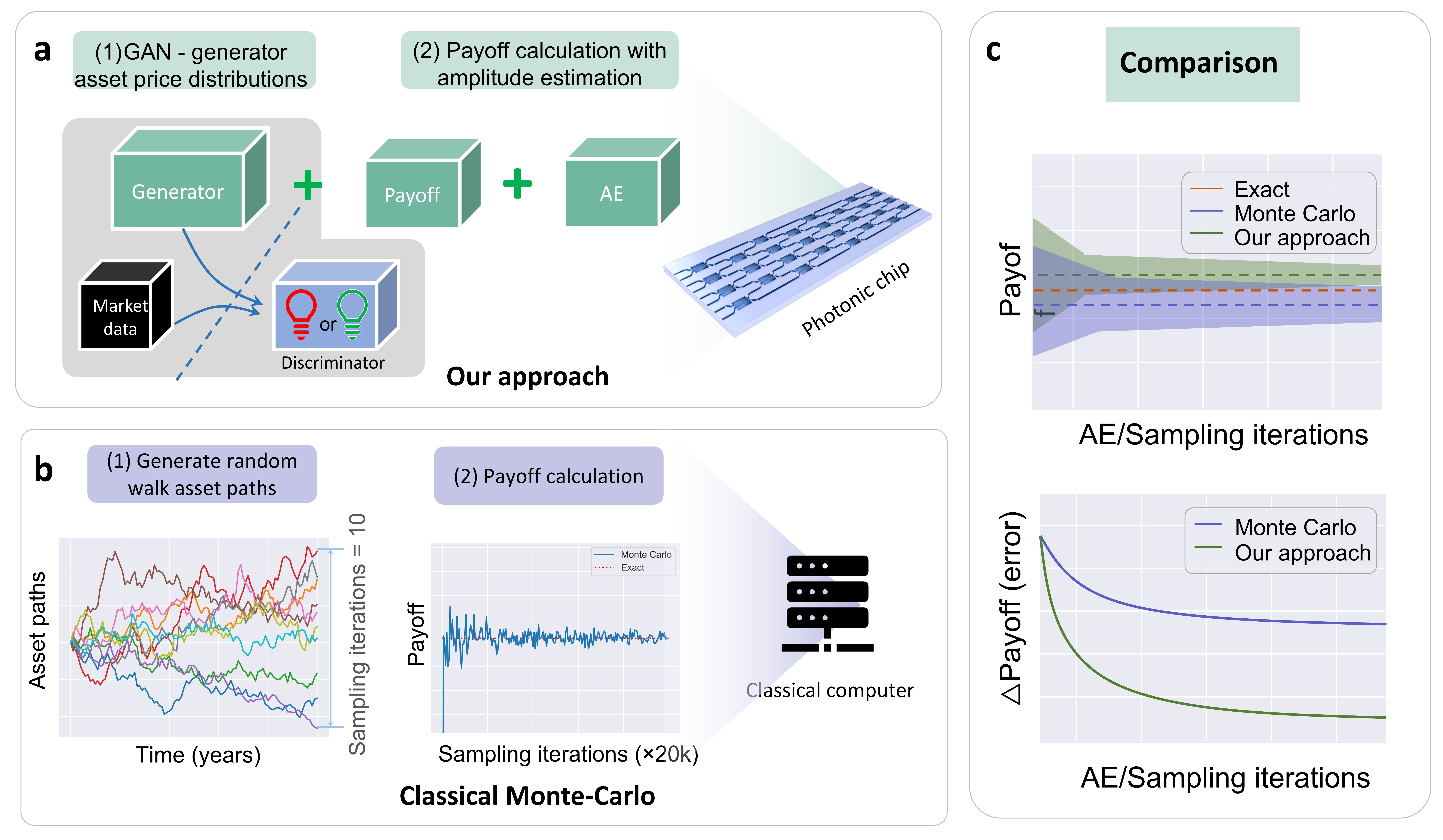}
\caption{\textbf{The schematic of the unary approach to option pricing, compared to the classical Monte-Carlo method.} 
\textbf{a}, The integrated photonic chip with the unary algorithm, consisting of a generator of the generative adversarial network (GAN), the payoff calculation, and the quantum amplitude estimation for acceleration.
\textbf{b}, Monte Carlo simulation on a classical computer, which first generates the future asset price paths based on random variables, and then calculates the return. The accuracy relies on extensive simulations of random walk asset paths.
\textbf{c}, Expected acceleration of the convergence of payoff errors, compared to classic Monte Carlo simulations. Shaded areas in the top inset indicate statistical uncertainty. }
\label{fig: intro}
\end{figure*}

Whereas, for specialized application tasks such as option pricing, there is no need to use universal quantum computers. Photonic circuits can provide fundamental functions that can be combined to implement specific algorithms~\cite{zhang2022resource,wang2023image,fu2023photonic,xu2021optical,liao2021all,zhou2022photonic,romero2021variational}, which would be practical and efficient for user-cased application scenarios. Moreover, the reduced energy costs of photonic computing have been a driving force behind works on dedicated photonic chips for machine learning, and algebra \cite{zhang2021optical,zhang2019artificialnanopho,spall2020fullyenergy,yan2022metasurface,fu2023photonic}. Therefore, we demonstrate a unary (against binary) approach in a photonic chip for option pricing. Compared to the binary approach, the unary approach~\cite{ramos2021quantum} has a remarkably simplified structure and depth of quantum circuits and is especially suitable for linear optical circuit realizations in photonic chips. The unary scheme also allows a post-selection strategy for error mitigation. Additionally, we demonstrate the generative adversarial learning to upload the probability distribution implicitly given by data samples into the photonic chip. Generative adversarial learning has previously only been demonstrated in superconducting and optoelectronics devices~\cite{zoufal2019quantum,hu2019quantum,huang2021experimental,huang2021quantum,wu2022harnessing,lloyd2018quantum} . Compared with traditional Monte Carlo methods, our approach shows high accuracy and significantly speeds up. It provides a promising avenue for interdisciplinary research in quantum machine learning and financial problems, paving the way for the development of practical photonic processors for quantitative financial applications. It can greatly improve the efficiency and quality of financial services, which is of great significance to the rapid and steady development of the financial industry.

\section{Chip design for unary option pricing}
In this work, we focus on European option pricing, and the expected payoff of options is given by
\begin{equation}
    C(S_T,K)=\int_K^\infty(S_T-K)dS_T
\end{equation}
where $S_T$ is the asset price at time $T$, and $K$ is the strike price. Figure~\ref{fig: intro} shows the overall scheme of operation of our photonic-chip-based unary approach. The photonic chip (Fig.~\ref{fig: intro}a) consists of a generative adversarial network (GAN) and an option pricing part that includes payoff computation and amplitude estimation. In contrast to the classical Monte-Carlo approach (Fig.~\ref{fig: intro}b) that requires huge computing power to simulate future asset prices to obtain an accurate solution, our approach is expected to show a speed-up in the convergence of the standard error of estimated payoff (Fig.~\ref{fig: intro}c), which is proved experimentally later in this section.

\begin{figure}[htb]
\centering
\includegraphics[width=0.49\textwidth]{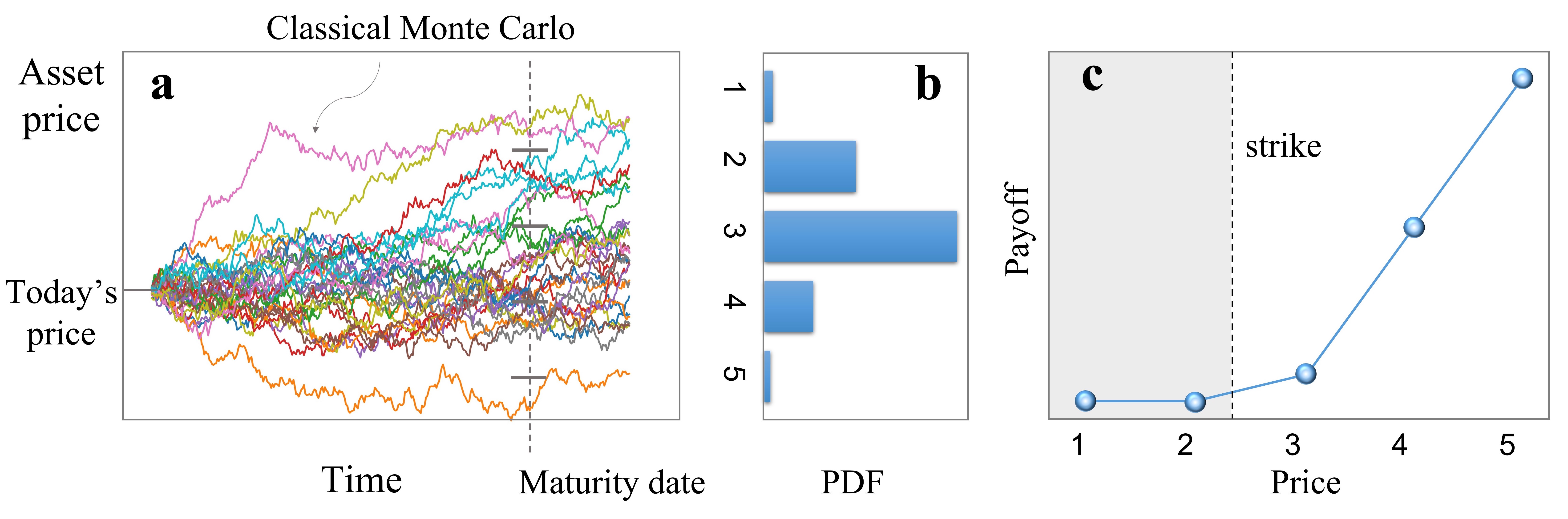}
\caption{\textbf{The mapping of asset prices to unary basis.} \textbf{a}, Classical Monte Carlo paths partitioned into different unary bases. \textbf{b}, The probability density function (PDF) according to the defined unary basis. \textbf{c}, The payoff value calculated according to the PDF and asset prices.}
\label{fig:option}
\end{figure}

\begin{figure*}[t]
\centering
\includegraphics[width=0.75\textwidth]{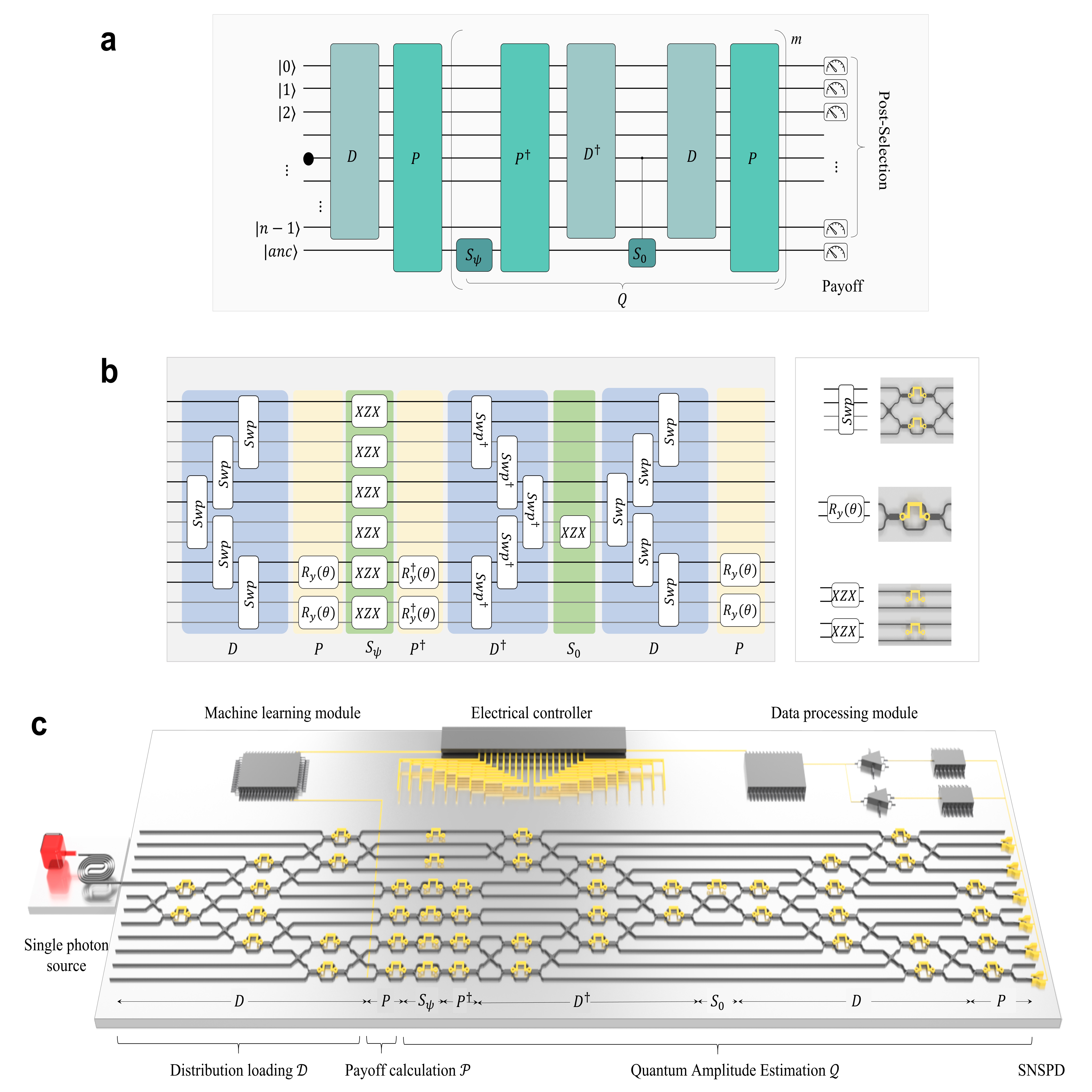}
\caption{\textbf{The photonic chip design for the unary option pricing algorithm.} 
\textbf{a}, The algorithmic model of unary option pricing. The input state consists of a $n$-dimensional qudit and a 2-dimensional ancilla. The following modules are contained, $\mathcal{D}$ - for distribution loading, $\mathcal{P}$ - for payoff calculation, and $\mathcal{Q}$ - the quantum operator for Amplitude Estimation. The amplification module $\mathcal{Q}$ is performed sequentially by $\mathcal{S}_{\psi}\rightarrow \mathcal{P}^\dagger\rightarrow \mathcal{D}^\dagger\rightarrow \mathcal{S}_0\rightarrow \mathcal{D}\rightarrow \mathcal{P}$. The expected payoff is obtained by measuring the ancilla. 
\textbf{b}, The optical circuit model by transforming the algorithmic model to linear optical operators. Each element of the unary basis is represented by two waveguides, extending the $n$-bin unary basis to a $2n$-dimensional Hilbert space. Relevant linear optical operators, $swp$, $R_y(\theta)$, and $XZX$ are listed with their waveguide structures. 
\textbf{c}, The photonic chip design and architecture. The chip is designed by transforming the optical path model into waveguide structures and realizes the distribution loading, payoff calculation, and amplitude estimation sequentially. The distribution loading is trained as a GAN embedded in the machine learning module.
}
\label{fig:unary-chip}
\end{figure*}

The unary approach to option pricing encodes an asset price distribution into the unary basis of a quantum register, as shown in Fig.~\ref{fig:option}. A binning scheme is applied such that Monte Carlo paths that would belong to the same interval of asset prices end up in the same bin. Each bin is then mapped to an element of the unary basis, whose coefficient is the ratio of the number of Monte Carlo paths in that bin to the total number. The accuracy of unary encoding is bounded by the number of bins that can be stored in a quantum state, \emph{i.e.}, the usable dimension of the high-dimensional unary state. Based on the unary basis, Figure~\ref{fig:unary-chip}a depicts the algorithmic model for unary option pricing, which consists of three modules: a distribution loading module $\mathcal{D}$ that loads the asset price distribution into a quantum state, a payoff calculation module $\mathcal{P}$ that computes the expected return, and a quantum amplitude estimation module $\mathcal{Q}$ to gain a quadratic speed-up over classical sampling to reach a target accuracy.

Figure~\ref{fig:unary-chip}b depicts the optical circuit model, whereby each module of the unary algorithm is mapped to a linear optical operator. We represent the high-dimensional state by path encoding a single photon using $n$ optical waveguides. The superposition of a single photon traveling through different waveguides directly encodes the unary basis. This high-dimensional state can be written as $\ket{\psi}=\sum_{i=0}^{n-1} \sqrt{p_i}\ket{i}$, where $p_i$ represents the probability of observing a photon in the waveguide mode $\ket{i}$, and these probabilities conform to $\sum_{i=0}^{n-1}p_i = 1$. The payoff calculation requires an ancilla qubit to store the expected return for each asset price, expanding the Hilbert space of the algorithm to 2$n$. To avoid non-local controlled gates in the photonic chip implementation, we instead add an ancillary waveguide to each of the $n$ unary waveguide modes to represent the effect of the ancilla qubit. Each element of the unary basis is now represented by two waveguides. This way, the controlled operations of the original algorithm are converted to linear transformations on the optical circuit. The architecture of the photonic processor with the detailed chip design is shown in Fig.~\ref{fig:unary-chip}c, which replaces each linear optical operator with the corresponding waveguide structure. The entire chip is reconfigurable via wire bonds and integrated thermo-optic phase shifters. 

In the \emph{\textbf{distribution loading module}} $\mathcal{D}$, a single photon is incident into the chip from a waveguide in the middle of the circuit, which encodes the ancilla in its $\ket{0}$ state, \emph{e.g.}, for a 3-asset case, the initial input state can be written as the tensor product of the middle unary qudit and the ancilla qubit as $[0,1,0]\otimes[1,0]=[0,0,1,0,0,0]$. The distribution of asset prices is then uploaded to the different waveguides using a linear depth circuit. This distribution loading circuit spreads the superposition to the neighboring basis using \emph{swp} operators
\begin{equation}
    swp = \begin{pmatrix}
    &I & & &\\
    & &\sqrt{p} &\sqrt{1-p} &\\
    & &\sqrt{1-p} &-\sqrt{p} &\\
    & & & &I\end{pmatrix}\otimes I,
\end{equation}
where $p$ depends on the target distribution. The procedure is repeated until the edge of the circuit is reached. The distribution loading module can be reconfigured to obtain any target probability distribution in the unary representation. Precisely, given $n$ assets, the depth of the circuit is always $\lfloor (n+1)/2 \rfloor$, and the loading of any known probability distribution onto the unary basis depends on $(n-1)$ splitting parameters $p$. The generator of a GAN is embedded in this module. The GAN is employed to capture the probability distribution underlying given market data. The details are presented in the next section.

The \emph{\textbf{payoff calculation module}} $\mathcal{P}$ encodes the expected payoff as the probability of measuring the photon in the waveguides encoding the ancilla in state $\ket{1}$, using rotation operations between the two waveguides of each element of the unary basis. The rotations encode the expected return for each asset price in the distribution. This action, labeled $\mathcal{P}$, can be written as
\begin{equation}
\small
    \mathcal{P} = \begin{pmatrix}
    &M_0 & & & \\
    & &M_1 & & \\
    & & &\ddots & \\
    & & & &M_{n-1}
    \end{pmatrix},\,M_i=\begin{pmatrix}
    &{\rm cos}\theta_i &{\rm -sin}\theta_i\\
    &{\rm sin}\theta_i &{\rm cos}\theta_i
    \end{pmatrix}
\end{equation}
for a $2n$-waveguide, $n$-bin example. 

A \emph{\textbf{quantum amplitude estimation module}} $\mathcal{Q}$ is applied to achieve a quantum speed-up. Various Amplitude Estimation techniques have been presented that are friendly to NISQ devices~\cite{suzuki2020amplitude, aaronson2020quantum, grinko2021iterative}. Here, we implement an amplitude estimation algorithm without quantum phase estimation in the photonic circuit, following the technique used in Ref.~\cite{ramos2021quantum}. Increasing steps of amplitude amplification are applied to estimate the relevant amplitudes with up to a square root advantage oversampling from the original distribution. This amplification module $\mathcal{Q}$ is performed by applying the following operators. First, $S_\psi$ identifies the amplitudes that encode the expected payoff and reverses their signs. Explicitly, for the 3-asset example at hand, such operation is $S_\psi=diag(1,-1,1,-1,1,-1)$, and is realized experimentally by applying a phase shift of $\pi$ on the second waveguide of each element of the unary basis. 
Then, the original operations are reversed, that is, the inverse of the payoff calculator $\mathcal{P}^\dagger$ and the distribution loading $\mathcal{D}^\dagger$ are applied.
An operator $S_0$ follows, which reverses the sign of the initial state of the computation.
Experimentally, it is applied by introducing a phase shift of $\pi$ to the waveguide where the photon was introduced.
The last step is to repeat the distribution loading $\mathcal{D}$ and the payoff calculator $\mathcal{P}$ modules. 
The amplitude amplification operator $\mathcal{Q}=\mathcal{P}\cdot\mathcal{D}\cdot\mathcal{S}_0\cdot\mathcal{D}^\dagger\cdot \mathcal{P}^\dagger\cdot\mathcal{S}_\psi$ is repeated a different number of times, and the results are processed to estimate the expected payoff. This technique provides up to a quadratic speedup over ordinary sampling in the number of calls to the $\mathcal{D}$ and $\mathcal{P}$ operators to reach the same confidence level, see theoretical derivations in Appendix C.

\begin{figure}[t]
\centering
\includegraphics[width=0.49\textwidth]{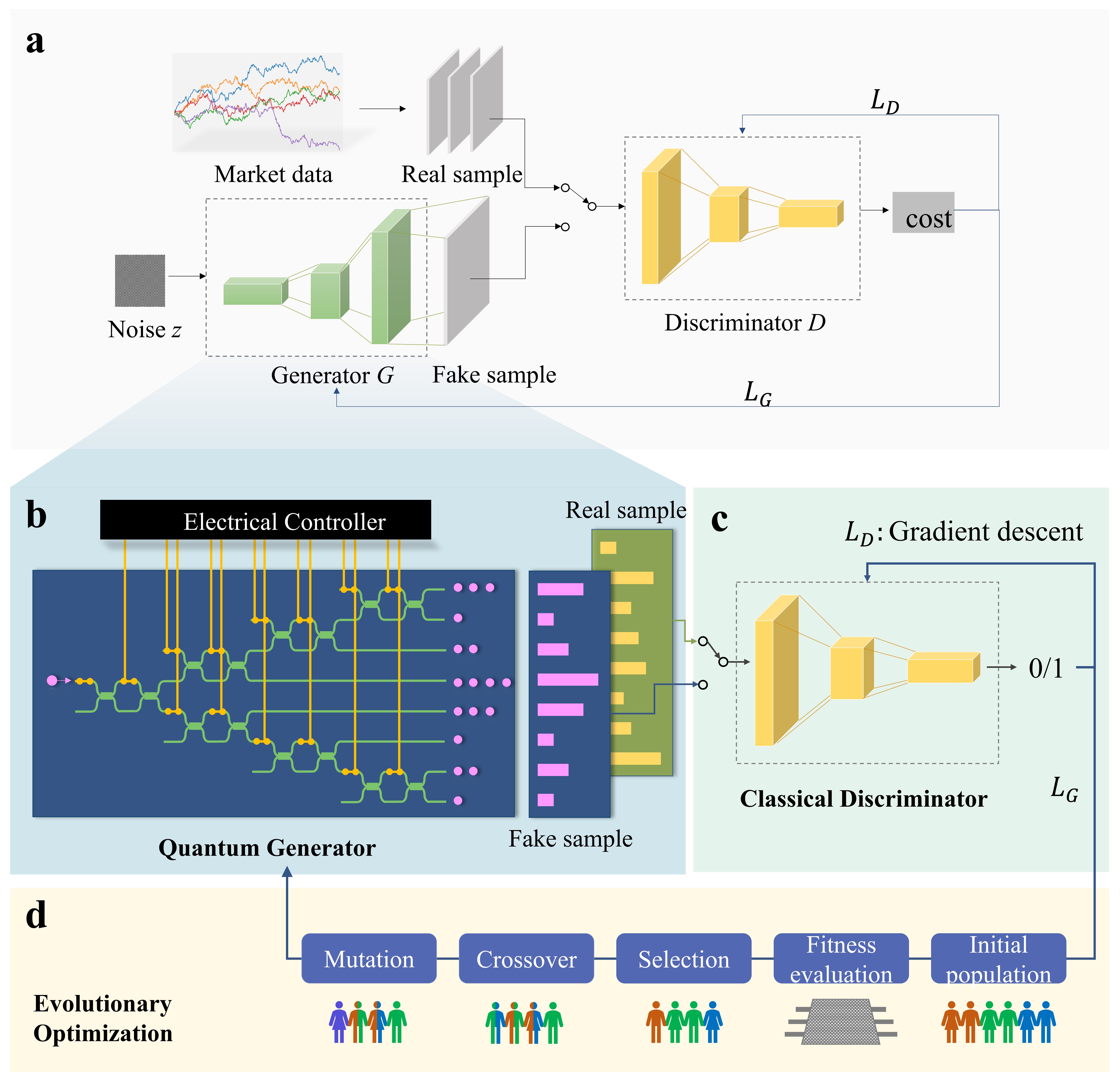}
\caption{\textbf{The GAN on the photonic chip for precise asset distribution uploading.}
\textbf{a}, The algorithm of GAN, which is composed of a generator and a discriminator.
\textbf{b}, The generator implemented by a variational photonic circuit, which is trained on-chip in real time. The probability distributions accumulated on the waveguide paths are used as fake samples. Real samples are the training targets taken from market data in real applications.
\textbf{c}, Classical discriminator consisting of sequential convolutional layers, and trained by a gradient descent algorithm. The discriminator aims to distinguish the source of the input sample, from the generator or a real distribution. The cost function is calculated from the discriminator output and used to train the discriminator itself and the generator.
\textbf{d}, The generator is trained by an evolutionary optimization procedure where populations (\emph{e.g.}, different configurations of the generator ansatz) are generated, evaluated, and iterated. The evaluation is accomplished using the scores granted by the discriminator. New generations are produced via the operators of selection, crossover, and mutation of current populations.
}
\label{fig:qgan-overview}
\end{figure}

\section{GAN for distribution uploading}
A Generative Adversarial Network~(GAN) is implemented in the distribution loading module with on-chip training for real-time noise perception. The goal of the GAN is to obtain an intelligent generator at the chip parameter level that captures the probability distribution behind the given market data without simulating enormous random paths, accumulating data statistics, and then fitting them into the chip architecture. With the GAN, we can efficiently load the classical data, \emph{i.e.}, the probability distribution underlying market data, into quantum states and obtain more precise payoff calculations with the presented unary option pricing methods.

GANs train a generator (G) to synthesize semantically meaningful data from standard signal distributions, as well as a discriminator (D) to distinguish \textit{real} samples in the training dataset from \textit{fake} ones produced by the generator~\cite{goodfellow2014generative}, as depicted in Fig.~\ref{fig:qgan-overview}a. As its adversary, the generator aims at deceiving the discriminator by producing more realistic samples. Training GANs involve the search for a Nash equilibrium of a two-player game between a generative and a discriminative network, which can be formulated as:
\begin{equation}
    \min_G\max_D\mathbb{E}_{x\sim p_{\rm real}}[{\rm log} (D_\phi(x))]+\mathbb{E}_{z\sim p_z}[{\rm log}(1-D_\phi(G_\theta(z)))],
\end{equation}
where the generative network $G_\theta$ takes noisy samples $z$ from a normal or uniform distribution $p_z$ as input, $x$ comes from the real distribution $p_{\rm real}$. The discriminative network $D_\phi$ tries to distinguish the generated (fake) sample $G_\theta(z)$ and the real sample $x$, by projecting their output to $\{0,1\}$. The $\theta$ and $\phi$ are the free parameters that construct the generator and the discriminator.
The training procedure is complete when the generator wins the adversarial game, that is, the discriminator cannot make a better decision than random guesses on the validity of a sample. 

We develop a hybrid GAN implementation that consists of a generator network in the photonic chip, a classical discriminator network, and a control system that communicates between the classical computer and photonic chip, all depicted in Fig.~\ref{fig:qgan-overview}b-d. The generator is parameterized by the angles on the phase shifters that are reconfigurable through the thermo-optic effect, induced by applying tiny electrical power to the integrated heaters. Instead of a noise distribution as input, we utilize the uncertainty of photons appearing at different waveguide modes, to achieve the equivalent randomness for the generator. The fake samples are the probability distribution of the photons at the different waveguide modes. The real samples are drawn from the desired probability distribution, a log-normal or normal distribution for the examples presented in Fig.~\ref{fig:wqgan-results}. The fake and real samples sequentially enter the classical discriminator to achieve the classification results. The discriminator is a classical neural network implemented with TensorFlow. We then explicitly discuss the training of the GAN with data samples drawn from log-normal distribution and normal distribution.

The training process of the GAN in a photonic chip introduces two challenges, the difficulty of obtaining gradients due to the stochastic nature of measurements, and the phenomenon that the discriminator easily overpowers the generator. To circumvent these problems, we propose a hybrid training strategy, where the generator is optimized under a gradient-free \textit{evolutionary} algorithm, while the classical discriminator uses a gradient descent optimizer. Additionally, the Wasserstein distance~\cite{arjovsky2017wasserstein,chakrabarti2019quantum} is used to train the GAN, which changes the dynamic between the generator and the discriminator. In this new GAN scheme, the discriminator acts as a \textit{critic}, instead of classifying. It aims to give a high score to real instances over fake ones, effectively alleviating the problem of unstable GAN training.

\section{Results and Discussions}

\begin{figure}[t]
\centering
\includegraphics[width=0.49\textwidth]{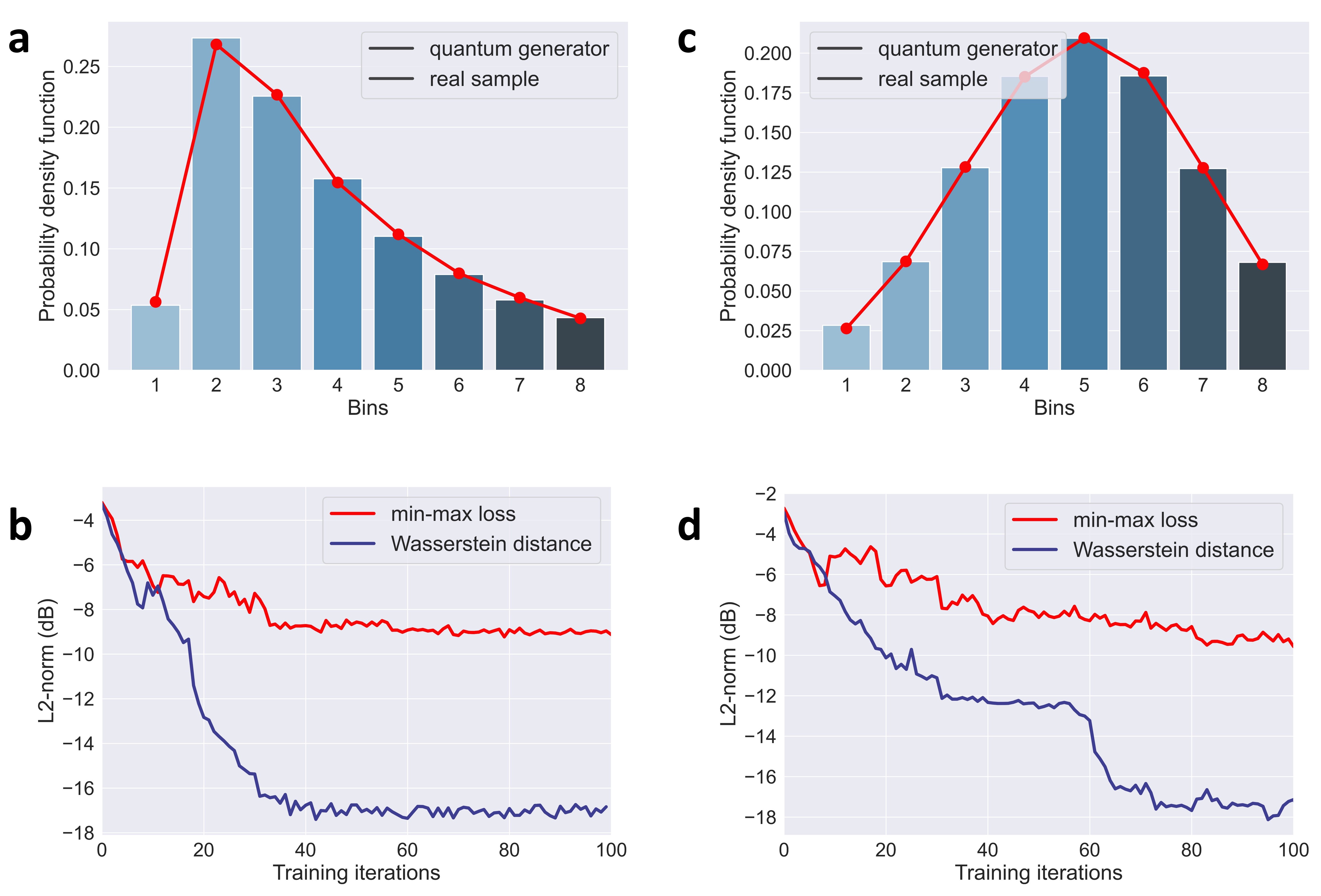}
\caption{\textbf{The experimental training performance of the GAN under Wasserstein distance.} \textbf{a}, \textbf{c}, Comparison between the probability distributions obtained experimentally from the generator (solid line with data points) and the target distribution (histogram). \textbf{b}, \textbf{d}, Evolution of the $\ell_2$ norm between the fake and real samples with increasing training iterations. \textbf{a}, \textbf{b}: Log-normal distribution; \textbf{c}, \textbf{d}: Normal distribution.}
\label{fig:wqgan-results}
\end{figure}

\textbf{GAN results.}
Our chip can accommodate the entire option pricing process of distribution loading, payoff calculation, and amplitude estimation, for 3 option assets. 
However, training 3 bins is too trivial to demonstrate the ability to implement GANs on a photonic chip. 
Here, to demonstrate the GAN, we employ a chip that supports up to 8 bins to demonstrate the generation of the probability distribution. Figure~\ref{fig:wqgan-results}a shows the probability distribution of the generator output compared to the real log-normal distribution. Figure~\ref{fig:wqgan-results}b shows the convergence of the $\ell_2$ norm between the fake and real samples, of 100 training iterations. For a generator output $\it \textbf{g}$ and a real distribution $\it \textbf{x}$, the $\ell_2$ norm is defined as 
\begin{equation}
    \ell_2 = \sqrt{\sum_{i=1}^m (x_i-g_i)^2}.
\end{equation}
The results for a target normal distribution are shown in Figs.~\ref{fig:wqgan-results}c and \ref{fig:wqgan-results}d. For both examples of lognormal and normal distribution, the final $\ell_2$ norm between the generator output and the real distribution stabilizes at -18 dB. 

By training this generative model directly on the photonic chip, we bypass the need to solve the BSM equations while capturing the nuances that the simplified method overlooks. Concurrently, it incorporates environmental elements that are hard to model, such as crosstalk and chip imperfections into the GAN training. Another feature of using GANs for the amplitude distribution step is that we can tailor the variational ansatz to construct short-depth circuits for a given degree of accuracy, even in a more general case with multiple photons.

\begin{figure*}[t]
\centering
\includegraphics[width=0.69\textwidth]{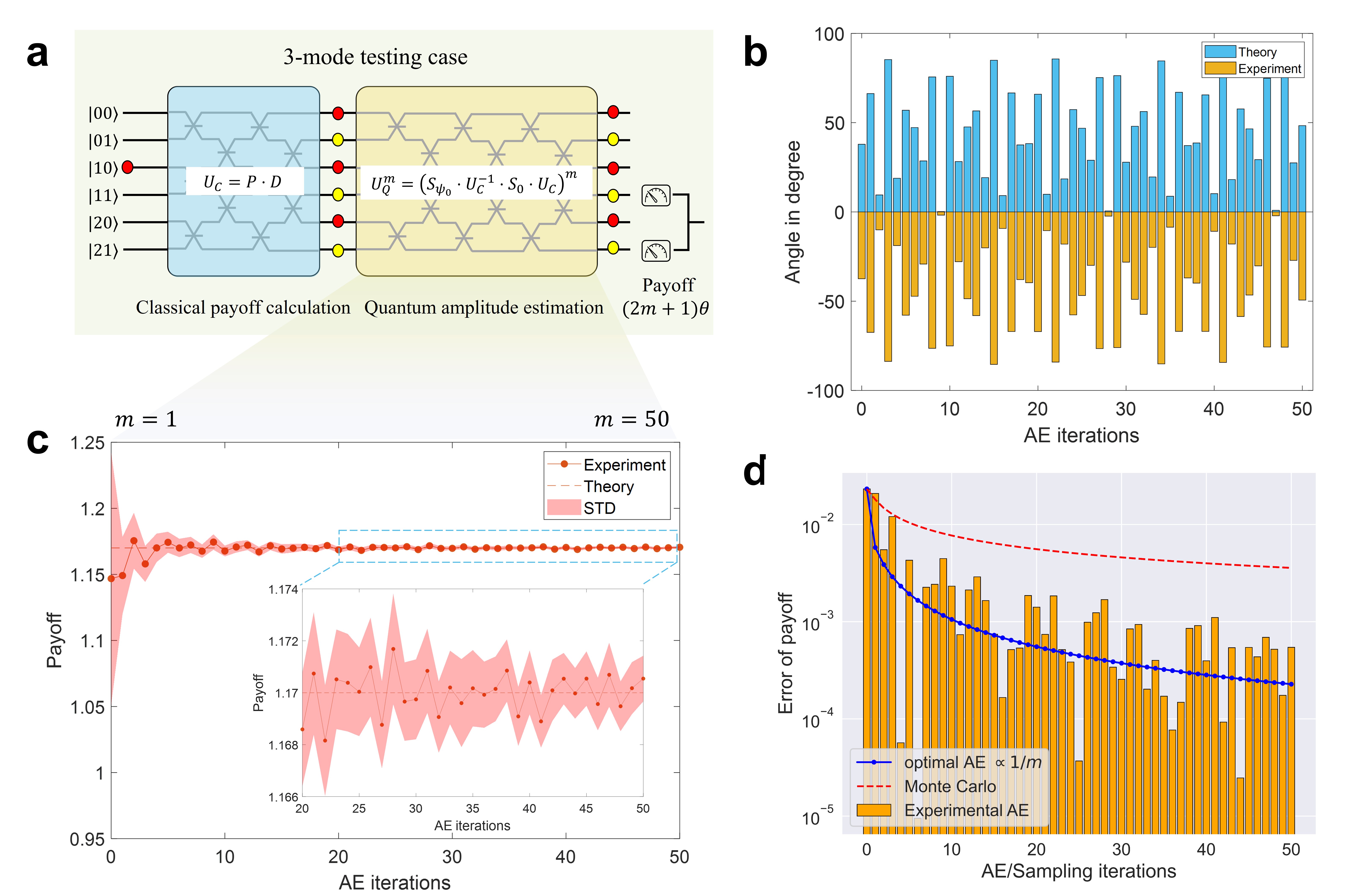}
\caption{\textbf{The experimental results of option pricing with three asset values.} \textbf{a}, Illustration of the optical chip with the payoff calculation and the amplitude estimation module. The operator $\mathcal{Q}$ is repeated up to $m$ ($m\leq50$) times. The payoff is measured on the waveguides that encode the ancilla in state $\ket{1}$ when the asset price is larger than the pre-defined strike value.
\textbf{b}, Comparison between the theoretical expectations and experimental results of the payoff, represented in angles. The raw angles $(2m+1)\theta$ are shifted back to the original angles $\theta$, and the differences from the theory expectations are recorded as errors.
\textbf{c}, Standard deviation (STD) of the expected payoff with increasing iterations of the amplitude estimation module. The STD converges from the initial $\sim0.2$ to less than 0.004. The iterations from 20 to 50 are highlighted. \textbf{d}, Error in payoff estimation between theoretical and experimental results, with increasing iterations of amplitude estimation. It shows a speed-up in convergence compared to the Monte Carlo method.}
\label{fig:payoff-results}
\end{figure*}

\vspace{0.5em}
\noindent\textbf{Unary option pricing results.} As a proof of principle, the fabricated photonic chip supports an option pricing problem with three asset values, whose schematic diagram is shown in Fig.~\ref{fig:payoff-results}a. The chip has 6 waveguide inputs: each pair represents one element of the unary basis and the ancilla qubit state. The chip is divided into distribution loading, payoff calculation, and then $m$ runs of amplitude amplification. To stay within the depth constraints of this proof-of-concept, the circuit's unitary matrix is multiplied and uploaded into the photonic chip at a constant depth. The single-photon measurement is performed at the waveguide modes that represent ancilla state $\ket{1}$ for asset prices larger than the strike value. The comparison between the theoretical payoff and the estimation achieved experimentally is shown in Fig.~\ref{fig:payoff-results}b, with increasing iterations of amplitude estimation. The performance of the amplitude estimation is shown in Figs.~\ref{fig:payoff-results}c and \ref{fig:payoff-results}d. In Fig.~\ref{fig:payoff-results}c, the dotted line represents the theoretical payoff expectation, the solid line with data points represents the experimental results, and the shaded area represents the standard deviation (std) of 50 measurements performed in each step of amplitude estimation. The progression of $m$ from 0 to 50 ($m=0$ being a classical sampling of the payoff calculation) demonstrates the convergence of the standard deviation. Similarly, in Fig.~\ref{fig:payoff-results}d, we visualize the convergence of the payoff error with more amplitude estimation runs. The amplitude estimation improves the accuracy of the expected payoff for a certain number of circuit runs.

The structure of the unary algorithm allows a simple but efficient design of photonic chips, especially when loading probability distributions into quantum registers since only local interactions between neighboring waveguides are required. This, however, is inaccessible for the binary alternative, where high connectivity is required to offset the exponential Hilbert space available. The optical circuit that implements the unary approach requires a linear number of waveguides scaling with the required precision, which coincides with the remarkable scalability of photonic chips. Instead, avoiding the use of controlled operations \textit{via} ancilla waveguides bypasses one of the main bottlenecks of photonic chips in quantum computing, the obstacle of realizing photon interactions.

Speedup is achieved in our work, much akin to proposals for quantum search without entanglement~\cite{lloyd1999entanglement,meyer2000sophisticated}, whereby a polynomial speedup of unstructured search is achieved with a single photon at the cost of exponential resources. In particular, isomorphisms exist between a system of $n$ qubits and a qudit residing in a $2^n$ Hilbert spaces (the systems presented in this thought experiment and our implementation)~\cite{ekert1998entanglement}, thereby the unary implementation on a photonic chip display entanglement in the path encoding of the single photon. Coherent light can achieve a similar effect, with a high sampling rate, which is advantageous in near-term use case scenarios, with some trade-offs for the random behavior of single photons in the generator part of the GAN.

The presented avenue to achieve speed-up in option pricing is scalable in the photonic chip. It transforms the unary algorithm's need for increasing qubits into a need for waveguide paths, which are highly scalable in photonic chips. For further scalability, in the presented experiment, the photon detectors placed in the ancilla waveguides could be combined into a single one, as only the counts of photons in any ancilla qubit are needed; hence significantly reducing the resources needed to scale this approach to meaningful problems. The energy efficiency of photonic chips also promises a relevant advantage beyond a complexity separation between quantum and classical algorithms. Given an energy budget instead of a shot budget, the photonic implementation of the unary approach to option pricing can yield a significant advantage in the number of operations performed. 

\section{Conclusion}

This work is the first demonstration of photonic chips for financial applications. As a proof-of-concept, we implement the unary option pricing algorithm in a photonic chip for European options, which includes the generation of the amplitude distribution of the asset value, the evaluation of expected return, and amplitude estimation. We prove the high accuracy in calculating the payoff function, as well as the effectiveness of amplitude estimation in reducing the number of evaluations to reach the same degree of accuracy when compared to classical sampling. The unary representation remarkably simplifies the structure and depth of quantum circuits in the linear optical circuit implementation. Such photonic devices could eventually be an eco-friendly alternative to electronic circuits. Furthermore, we demonstrate an on-chip training of a GAN that successfully captures important market dynamics in real-life scenarios, bypassing the simplified assumptions in the BSM model that limit its accuracy, as well as the computational burden in solving the differential equations. Most importantly, the photonic chip could be potentially employed for other options pricing, paving the way for developing dedicated processors in finance applications.

\section*{Acknowledgements}

H.Z., L.X.W., J.I.L., and A.Q.L. jointly conceived the idea. H.Z., L.X.W., and H.C. designed the chip and built the experimental setup. F.G., G.Q.L., and X.S.L. fabricated the silicon photonic chip. H.Z., L.X.W., and Y.C.Z. performed the experiments. W.K.M., S.R.C., L.C.K., and J.I.L. assisted with the theory. All authors contributed to the discussion of experimental results. L.C.K., J.I.L., and A.Q.L. supervised and coordinated all the work. H.Z., S.R.C., J.I.L., and A.Q.L. wrote the manuscript with contributions from all co-authors.

\textbf{Funding.} These research works are supported by the Singapore Ministry of Education Tier 3 grant (MOE2017-T3-1-001), National Research Foundation grant (NRF2022-QEP2-02-P16), and the Start-up Fund of The Hong Kong Polytechnic University (P0046236).

\textbf{Data availability.} Data underlying the results presented in this paper are not publicly available at this time but may be obtained from the authors upon reasonable request.

\bibliography{sample}

\newpage

\section*{Appendix A: Experimental setup and single-photon generation.} 
\begin{figure}[hbt]
\centering
\includegraphics[width=0.32\textwidth]{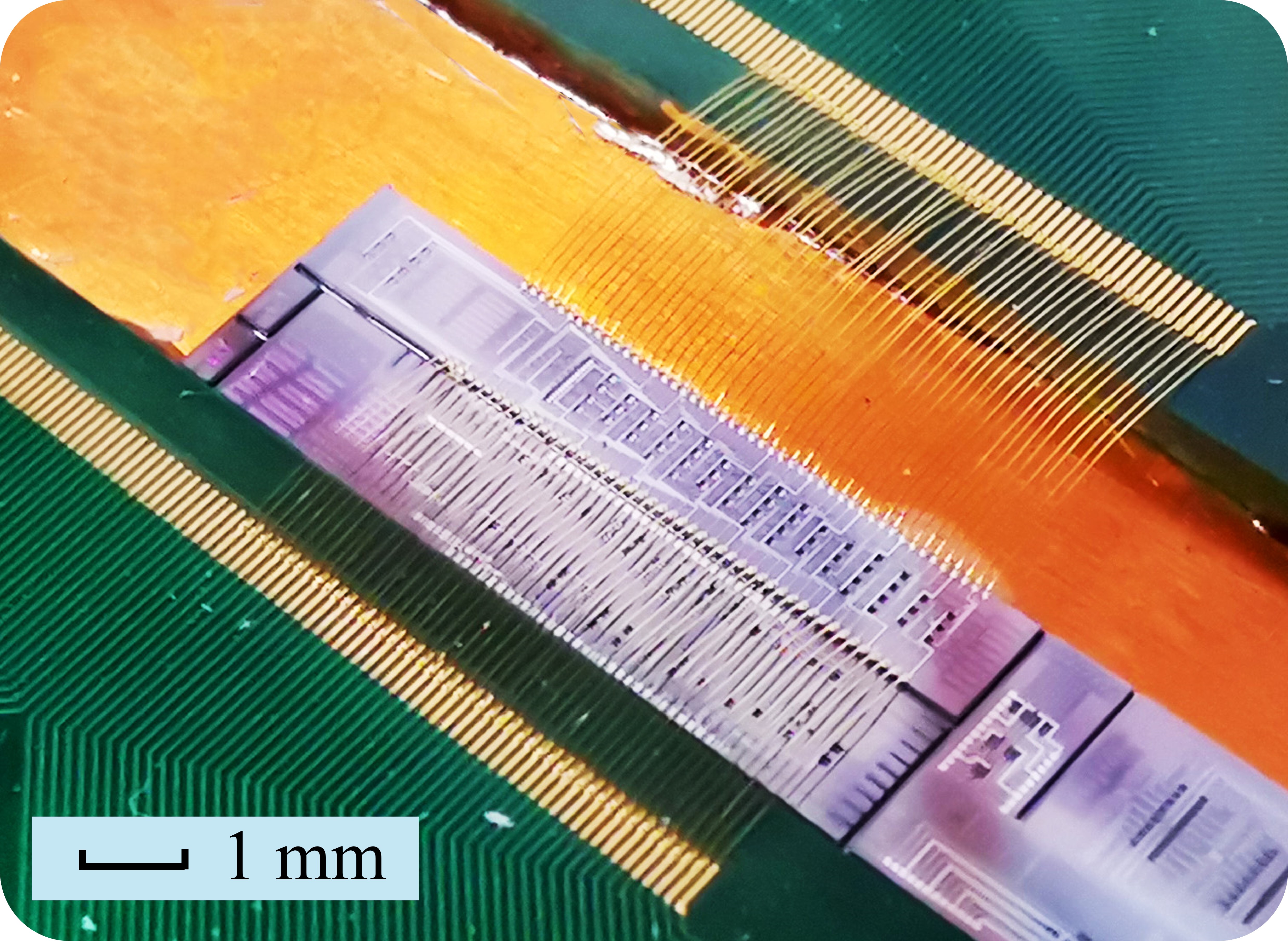}\\
\caption{\textbf{The fabricated quantum photonic chip.}}
\label{fig:chip}
\end{figure}

The entire packaged chip is shown in Fig.~\ref{fig:chip}. Each phase shifter is independently controlled by an electronic current driver with 1-kHz frequency and 12-bit resolution. Output photons are filtered via WDM to remove the residual photons, and then detected by superconducting nanowire single-photon detectors (SNSPDs) (from PhotonSpot, 100 Hz dark counts, 85\% efficiency). Polarisation controllers are placed before the SNSPDs as the detectors are polarization-sensitive. Time tagger (from Swabian Instrument) is used to count the single-photon events, which can support more than 40 million events per second. A temperature controller is used to stabilize the chip temperature and reduce thermal fluctuations caused by possible crosstalk.

Degenerated photon pair is used in our experiment. The pump laser is generated from the Ultrafast Optical Clock device (PriTel) with a repetition rate of 500 MHz, a central wavelength of 1550.116 nm, and a bandwidth of 1.9 nm. A dual pump scheme is employed to generate pairs of identical photons on chip with degenerated Spontaneous Four-Wave Mixing (SFWM) process. On the chip, the desired state $\ket{\psi}=\ket{11}$ is generated out of the 2 photon N00N state $\ket{\psi}=\frac{1}{\sqrt{2}}(\ket{20}+\ket{02})$, by configuring the phase value $\theta=\pi/2$ when interfering the 2 photons. 

\section*{Appendix B: European option pricing model.}
The Black-Scholes model is a typical economic model used to calculate the evolution of asset prices in financial markets, known as the European-option pricing problem. In this model, the evolution of option price $S_T$ at time $T$ is decided by two market properties, the interest rate $r$ and the volatility $\sigma$, which are expressed by a stochastic differential equation
\begin{equation}
    dS_T=S_T rdT+S_T \sigma dW_T,
\end{equation}
where $W_T$ describes a Brownian process, which is a continuous stochastic evolution starting at $W_0=0$ and consists of independent Gaussian increments. Specifically, let $\mathcal{N}(\mu,\sigma_s)$ be a normal distribution with mean $\mu$ and standard deviation $\sigma_s$, then the increment of two steps of the Brownian processes is $W_T-W_S\sim \mathcal{N}(0, T-S)$, for $T>S$. The stochastic differential equation can be approximately resolved to first order, and the solution is
\begin{equation}
    S_T = S_0 e^{(r-\frac{\sigma^2}{2})T} e^{\sigma W_T} \sim e^{\mathcal{N}((r-\frac{\sigma^2}{2})T,\sigma\sqrt{T})},
\end{equation}
which is a log-normal distribution. The process of solving the stochastic differential equation is valid for the simplified European option model, while for more practical cases, an analytical solution does not exist and even numerical simulation is costly. To get the expected return, a payoff calculation block is integrated over the resulting probability distribution. The payoff function is given by
\begin{equation}
    f(S_T, K) = max(0, S_T-K),
\end{equation}
producing an expected payoff
\begin{equation}
    C(S_T, K) = \int_K^{\infty}(S_T-K)dS_T,
\end{equation}
where $K$ is the strike. 

\section*{Appendix C: The theory of unary option pricing.}
By solving the aforementioned BSM model, the probability density function of the option price can be described by a log-normal distribution. We map this continuous price distribution into $n$ discrete values, which are the amplitudes of $n$ orthogonal quantum state basis, by using a probability loading operator $D$ acting on an initial state $\ket{\psi_{ini}}$ as
\begin{equation}
    D\ket{\psi_{ini}} = \sum_{i=0}^{n-1}\sqrt{p_i}\ket{\psi_i}_n,
\end{equation}
where each state $\ket{\psi_i}$ represents a discrete option price value $S_i$, and $p_i$ is the corresponding probability. These quantum state bases are orthogonal so that $\bra{\psi_i}\ket{\psi_j} = \delta_{ij}$. The payoff is obtained by accumulating the asset value under its corresponding probability. The payoff of the European option in this discrete scenario can be simplified as 
\begin{equation}
    C(S_T,K) = \sum_{0}^{n-1}p_i\cdot f(S_i,K) = \sum_{S_i>K}^{n-1}p_i\cdot(S_i-K), \\
\end{equation}
where $K$ is the strike price. The rotation angles after being normalized by the maximum asset price $s_{max}$ is given by
\begin{equation}
    \theta_i = \text{max}(0, {\rm arcsin}(\sqrt{\frac{s_i-K}{s_{max}-K}})).
\end{equation}
This payoff calculation can be mapped to the quantum model by introducing an ancilla qubit into the original quantum state followed by a controlled rotation gate $CR$ defined as 
\begin{equation}
    CR = \sum_{i=0}^{n-1}\ket{\psi_i}\bra{\psi_i}\otimes R_y(2\theta_i).
\end{equation}
\begin{figure}[tb]
\centering
\includegraphics[width=0.32\textwidth]{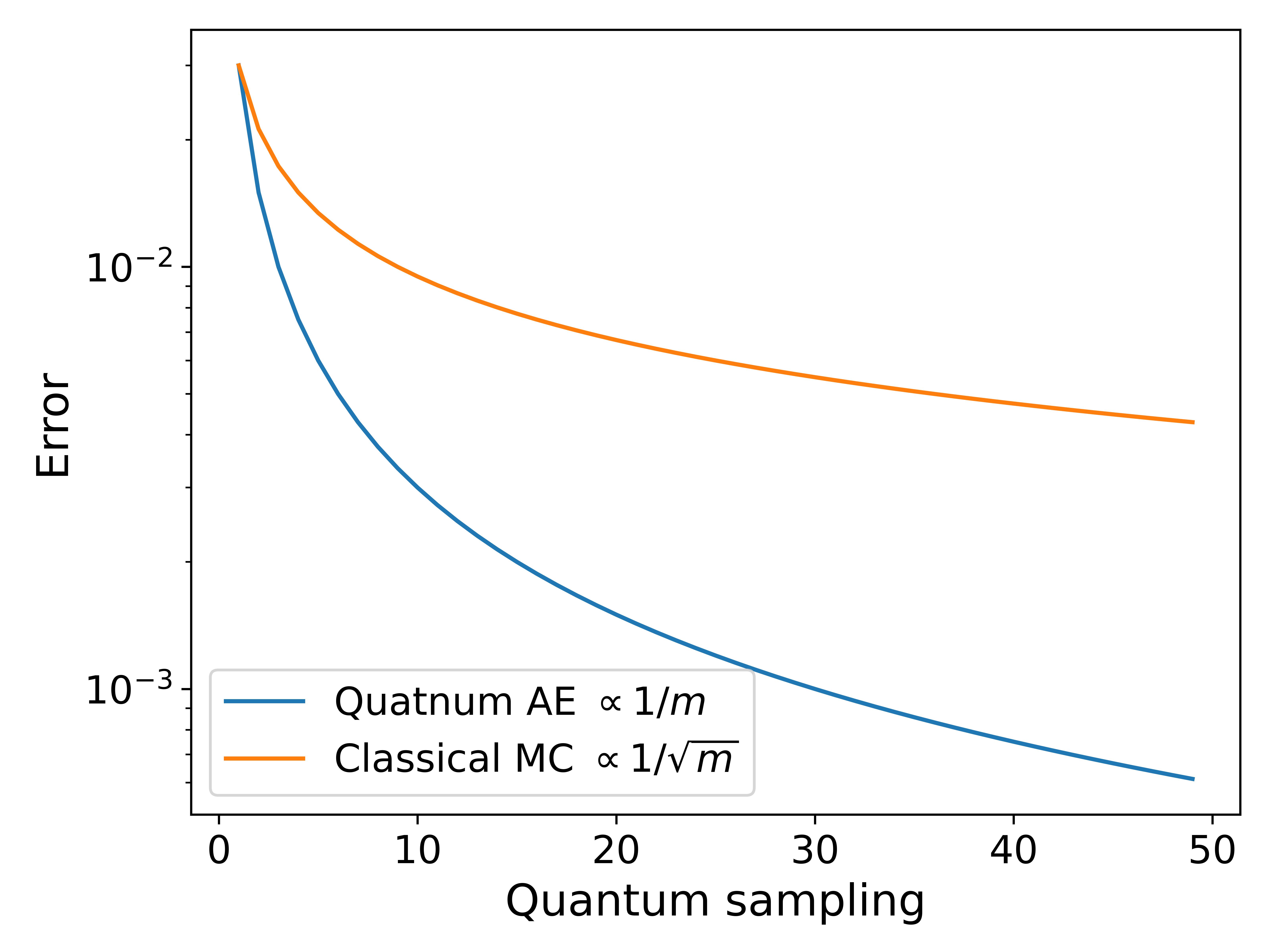}\\
\caption{\textbf{The simulation of the scaling of quantum AE and classical MC.}}
\label{fig:scaling}
\end{figure}
\begin{table*}[t]
\small
\centering
\caption{\textbf{Unary and Binary comparison.}}
\label{tab:1}
\begin{tabular}{p{4.2cm}p{4cm}p{4cm}}
\hline
Aspect                                                             & Unary approach & Binary approach \\ \hline
Representation                                                     & Intuitive - Single symbol repeated multiple times               &  Compact - Base-2 system with 0 and 1 symbols               \\
Chip architecture                                                  &   Simple - First-nearest neighbour connectivity             &  Complex - Full connectivity               \\
Amplitude estimation                                               &     Without phase estimation – feasible in linear optical circuits           & Require phase estimation - not feasible in linear optical circuits                \\
Gate count                                                         &  Linear (advantageous for near-term devices with $\textless$ 100 qubits)              &  Logarithmic, requires Toffoli gate               \\
Distribution loading error due to single-qubit error &  KL divergence of $10^{-3}$, one order of magnitude lower               &     KL divergence of $10^{-2}$            \\
Payoff deviation due to single-qubit error                         &  $\sim$ 25\%, 10\% more robust             & $\sim$35\%                \\ \hline
\end{tabular}
\end{table*}
Then, the expected payoff of the option price is related to the amplitude of the ancilla qubit in the form of 
\begin{equation}
\begin{aligned}
    \ket{\psi} &= CR\cdot\sum_{i=0}^{n-1}\sqrt{p_i}\ket{\psi_i}\otimes\ket{0}\\
    &= \sum_{i=0}^{n-1}\sqrt{p_i}{\rm cos}\theta_i\ket{\psi_i}\ket{0}+\sqrt{p_i}{\rm sin}_{\theta_i}\ket{\psi_i}\ket{1}.
\end{aligned}
\label{eq:s1}
\end{equation}
By measuring the ancilla qubit under basis $\ket{1}$, we can achieve the result as
\begin{equation}
    |\bra{1}\ket{\psi}|^2 = \sum_{i=0}^{n-1} p_i\cdot {\rm sin}^2 \theta_i = \frac{C(S_T,K)}{S_{max}-K}.
\end{equation}
Thus, the payoff of the option price can be directly read out from the measurement results of ancilla qubit under basis $\ket{1}$. Then we explain how the amplitude estimation works. The payoff calculation (Eq.~\ref{eq:s1}) can be simplified as
\begin{equation}
    CR\cdot D\cdot\ket{\psi_{ini}}\ket{0} = {\rm cos}\alpha \ket{\psi_a}\ket{0}+{\rm sin}\alpha \ket{\psi_b}\ket{1},
\end{equation}
where $\alpha$ is the normalized parameter, $\ket{\psi_a}$ and $\ket{\psi_b}$ are the normalized state
\begin{equation}
    \ket{\psi_a}=\sum_{i=1}^{n-1}\sqrt{p_i}{\rm cos}\theta_i\ket{\psi_i},~
    \ket{\psi_b}=\sum_{i=1}^{n-1}\sqrt{p_i}{\rm sin}\theta_i\ket{\psi_i}.
\end{equation}
The ancilla qubit is functioning as an indicator to identify the useful state. The amplitude amplification step begins by applying an oracle operator $S_\psi$ on the state $\psi$ with the form
\begin{equation}
    S_{\psi} = I - 2\sum_{i=0}^{n-1}\ket{\psi_i}\bra{\psi_i}\otimes \ket{0}\bra{0}
\end{equation}
to produce a sign change on the ancilla qubit state $\ket{0}$ that we want to perform the amplitude estimation. Then we add an inversion operation of the previous payoff calculation $CR$ and distribution loading operator $D$ followed by another sign flip operation $S_0$ on the initial state as
\begin{equation}
    S_0 = I-2\ket{\psi_{ini}}\bra{\psi_{ini}}\otimes\ket{0}\bra{0}
\end{equation}
and the last step is to apply $D$ and $CR$ again so that the amplitude estimation operator $Q$ can be written as 
\begin{equation}
    Q = CR\cdot D\cdot S_0\cdot D^\dagger\cdot CR^\dagger\cdot S_{\psi}
\end{equation}
By repeating the $Q$ operator m times, the full amplitude estimation can be represented as
\begin{equation}
    Q^{m}\cdot CR\cdot D\cdot\ket{\psi_{ini}}\ket{0}={\rm cos}(2m+1)\alpha\ket{\psi_a}\ket{0}+{\rm sin}(2m+1)\alpha\ket{\psi_b}\ket{1}
\end{equation}

Therefore, the measurement of ancilla qubit under basis $\ket{1}$ after repeated amplitude estimation would yield the results of ${\rm sin}^2(2m+1)\alpha$ for us to infer the payoff of the option price with improved accuracy. 
The amplitude estimation scheme we use here is an iterative approach~\cite{ramos2021quantum}. This procedure is based on the theory of confidence intervals for binomial distributions~\cite{wallis2013binomial} and uses samples of increasing Amplitude Amplification~\cite{brassard2002quantum} steps to better estimate the value of the target amplitude. 
The quantum amplitude estimation algorithm achieves a quadratic speedup when compared to classical Monte Carlo methods of pricing options,
\begin{equation}
    \mathcal{O}(\frac{1}{\sqrt{m}})\rightarrow\mathcal{O}(\frac{1}{m}).
\end{equation}
Where $m$ is the number of quantum samples used. The comparison between quantum and classical scaling factors is depicted in Fig.~\ref{fig:scaling}, exhibiting a trend that aligns well with our experimental results shown in Fig.~\ref{fig:payoff-results}d.

\section*{Appendix D: Unary and Binary comparison.}
The utilization of a unary approach~\cite{ramos2021quantum}, instead of the commonly-adopted binary approach~\cite{rebentrost2018quantum,stamatopoulos2020option} distinguishes this work from other quantum approaches. The key advantage of the unary method is its ability to implement all the necessary quantum operations within the option pricing algorithm using a linear optical circuit. In contrast, the binary approach relies on two-qubit controlled operations, which cannot be deterministically achieved in a photonic chip. The comparison between the unary and binary approaches is summarized in Table~\ref{tab:1}. The unary method has a simple chip architecture, no need for phase estimation, scalable gate count, accurate distributed loading, and robustness in payoff computation.

\end{document}